\documentclass[10pt, paper=a4, UKenglish]{article}
\usepackage{graphicx}
\def\Title#1{\begin{center} {\Large #1 } \end{center}}
\def\Author#1{\begin{center}{ \sc #1} \end{center}}

\newcommand\pubblock{\rightline{\begin{tabular}{l} Proceedings of the CTD 2023\\ \pubnumber\\
         \pubdate  \end{tabular}}}

\newenvironment{Abstract}{\begin{quotation} \begin{center} 
             \large ABSTRACT \end{center}\bigskip 
      \begin{center}\begin{large}}{\end{large}\end{center} \end{quotation}}

\newenvironment{Presented}{\begin{quotation} \begin{center} 
             PRESENTED AT\end{center}\bigskip 
      \begin{center}\begin{large}}{\end{large}\end{center} \end{quotation}}

\def\Acknowledgements{\bigskip  \bigskip \begin{center} \begin{large}
      \bf ACKNOWLEDGEMENTS \end{large}\end{center}}

\def\beq{\begin{equation}}
\def\eeq#1{\label{#1}\end{equation}}
\def\eeqn{\end{equation}}

\def\beqa{\begin{eqnarray}}
\def\eeqa#1{\label{#1}\end{eqnarray}}
\def\eeqan{\end{eqnarray}}

\let\bar=\overbar

\def\Dslash{\not{\hbox{\kern-4pt $D$}}}
\def\dslash{\not{\hbox{\kern-2pt $\del$}}}

\def\msb{{\bar{\ssstyle M \kern -1pt S}}}

\usepackage{color}
\usepackage{lineno}
\usepackage{subfig}
\usepackage{hyperref}

\newcommand\pubnumber{PROC-CTD2023-28}

\newcommand\pubdate{\today}

\newcommand{\conference}{Connecting the Dots Workshop (CTD 2023)\\
October 10-13, 2023}

\usepackage{fancyhdr}
\definecolor{mygrey}{RGB}{105,105,105}
\usepackage{amsmath}
\usepackage{amsfonts}
\usepackage{csquotes}
\usepackage{mathtools} 

\usepackage[maxcitenames=10, giveninits=true, maxbibnames=10, backend=biber, style=numeric-comp, sorting=none]{biblatex}
\bibliography{main}

\newcommand{\pt}{p_\text{T}}    
\newcommand{\effDmPt}[1]{\epsilon^\text{DM}_{\pt>#1}}
\newcommand{\effPerfectPt}[1]{\epsilon^\text{perfect}_{\pt>#1}}
\newcommand{\effLhcPt}[1]{\epsilon^\text{LHC}_{\pt>#1}}

\newcommand{\relu}{\operatorname{ReLU}}

\begin{document}


\large
\begin{titlepage}
\pubblock

\vfill
\Title{ High Pileup Particle Tracking with Object Condensation }
\vfill

\Author{Kilian Lieret\textsuperscript{1,2}, Gage DeZoort\textsuperscript{1}, Devdoot Chatterjee\textsuperscript{3}, \\Jian Park\textsuperscript{4}, Siqi Miao\textsuperscript{5}, Pan Li\textsuperscript{5}}
\vspace{0.3cm}
\begin{center}
\textsuperscript{1}Princeton University\\
\textsuperscript{2}Institute for Research and Innovation in Software for High Energy Physics (IRIS-HEP)\\
\textsuperscript{3}Delhi Technological University\\
\textsuperscript{4}University of Chicago\\
\textsuperscript{5}Georgia Tech
\end{center}
\vfill

\begin{Abstract}
Recent work has demonstrated that graph neural networks (GNNs) can match the performance of traditional algorithms for charged particle tracking while improving scalability to meet the computing challenges posed by the HL-LHC. Most GNN tracking algorithms are based on edge classification and identify tracks as connected components from an initial graph containing spurious connections. In this talk, we consider an alternative based on object condensation (OC), a multi-objective learning framework designed to cluster points (hits) belonging to an arbitrary number of objects (tracks) and regress the properties of each object. Building on our previous results, we present a streamlined model and show progress toward a one-shot OC tracking algorithm in a high-pileup environment.
\end{Abstract}

\vfill

\begin{Presented}
\conference
\end{Presented}
\vfill
\end{titlepage}
\def\thefootnote{\fnsymbol{footnote}}
\setcounter{footnote}{0}
%

\normalsize 


\section{Introduction}
\label{intro}
Traditional charged particle tracking algorithms at the Large Hadron Collider (LHC) are based on the combinatorial Kalman filter. 
However, this class of algorithms exhibits sub-optimal scaling with respect to pileup, rendering tracking a bottleneck for future experiments such as the High Luminosity LHC (HL-LHC)~\cite{Cerati_2016}.
This has prompted research into tracking algorithms leveraging graph neural networks (GNNs) or similar machine learning (ML) architectures demonstrating improved computational scaling. Recent results have confirmed that GNN-based algorithms can indeed achieve linear scaling with pileup~\cite{ju_exatrkx_2021,lazar_accelerating_2023}.

The majority of GNN approaches adopt an edge classification (EC) approach to tackle the tracking problem. 
Given an initial graph that connects all hits that potentially belong to the same particle, a GNN is trained to remove edges that connect hits belonging to different particles. 
Tracks can then be identified as connected components of the graph\footnote{This is simplified: Most state-of-the-art EC approaches use iterative \enquote{graph walking} algorithms to determine tracks based on EC scores rather than simply applying a threshold and using connected components.}, and subsequent steps assess track quality or fit track parameters.

This work considers an alternative approach based on clustering track hits in a learned latent space. Specifically, we employ object condensation (OC)~\cite{Kieseler_condensation_2020}, a multi-objective training scheme designed to cluster points (hits) matched to an arbitrary number of objects (tracks) and regress the properties of each object. 
Besides a general need for broad exploration of ML architectures in tracking, this class of approaches is motivated by several observations:
\begin{enumerate}
    \item 
    Most state-of-the-art EC approaches use a learned embedding space to build the initial graph edges between clusters of hits (metric learning, see Ref.~\cite{ju_exatrkx_2021}). 
    In an EC approach, this initial embedding space is discarded, whereas OC algorithms learn to refine it.
    \item 
    Edges in the EC approach serve two competing purposes: 1) facilitating message passing across the graph and 2) representing individual tracks in the graph.
    Though having more edges might facilitate a broader scope of message passing, the ultimate goal of an EC algorithm is to produce few edges, representing only real particle trajectories. 
    In contrast, edges are \emph{only} used for message passing in the OC approach, and tracks are constructed based on the node embedding space.

    This has important consequences: for example, any missing edge in an EC approach degrades performance (a track that is \enquote{broken up} cannot be fixed).
    However, we have shown~\cite{oc_chep_proceeding} that OC is not subject to the same limitation and can (to some extent) deal with missing edges. 
    \item 
    Because the OC approach directly addresses the relationship between hits and tracks, it can be trained to regress track properties; therefore, it is more suitable for one-shot architectures with little-to-no post-processing. 
    Training to regress track properties such as $\pt$ might also increase model robustness by imparting track physics onto the model.

\end{enumerate}

In this paper, we build on our previous results and show a new streamlined model that outperforms our previous architecture.
Instead of relying on a geometric graph construction, this model uses the learned clustering strategy of~\cite{ju_exatrkx_2021}.

We also discuss several other new insights and ongoing developments: The use of the Modified Differential Multiplier Method to balance different training objectives in object condensation, ongoing development of a model with GravNet layers, and ongoing development of a different approach using sparse transformers instead of GNNs.

\section{Dataset and metrics}
\label{sec:dataset}
\label{sec:metrics}
All results are produced using the TrackML dataset~\cite{amrouche_tracking_2020,amrouche_tracking_2021} that simulates a generic tracking detector geometry in the worst-case HL-LHC pileup conditions ($\langle\mu\rangle=200$).
We only consider the pixel detector layers (4 barrel layers and 7 layers in each endcap).
Each tracker event is represented as a graph by embedding track hits as nodes; a detailed summary of the features associated with each node is available in Ref.~\cite{oc_chep_proceeding}.

We use several different metrics to quantify the performance of our pipeline:
\begin{itemize}
    \item \textbf{Perfect match efficiency} ($\epsilon^\text{perfect}$): The number of reconstructed tracks that include all hits of the matched particle and no other hits, normalized to the number of particles.
    \item \textbf{LHC-style match efficiency} ($\epsilon^\text{LHC}$): The fraction of reconstructed tracks in which 75\% of the hits belong to the same particle, normalized to the number of reconstructed tracks.
    \item \textbf{Double majority match efficiency} ($\epsilon^\text{DM}$): The fraction of reconstructed tracks in which at least 50\% of the hits belong to one particle and this particle has less than 50\% of its hits outside of the reconstructed track, normalized to the number of particles.
    \item \textbf{Double majority match fake rate} ($f^\text{DM}$): The fraction of reconstructed tracks that does not satisfy the double majority match criterion, normalized to the number of reconstructed tracks.
\end{itemize}
Throughout most of the paper, we consider these metrics for \emph{particles of interest}, that is, particles with $\pt>0.9\,\mathrm{GeV}$, $|\eta|<4.0$ that have at least three hits. 
These metrics are denoted as $\effDmPt{0.9}$, $\effPerfectPt{0.9}$, $\effLhcPt{0.9}$, and $f^\text{DM}_{\pt>0.9}$.
For a more verbose definition of these metrics, see Ref.~\cite{oc_chep_proceeding}.
\section{A streamlined OC architecture}
\label{sec:pipeline2.0}
The pipeline outlined in this section comprises three stages: graph construction in a learned clustering space, object condensation, and postprocessing/track rendering.
\subsection{Graph construction}
The main improvement compared to the pipeline presented in Ref.~\cite{oc_chep_proceeding} is the use of a learned clustering approach to graph construction (GC).
This approach is very similar to that used in the Exa.TrkX EC pipeline in Ref.~\cite{ju_exatrkx_2021}.

We use a fully connected neural network (FCNN) to produce learned clustering coordinates for each track hit.
The FCNN takes the node features $z^{(0)}_i\in \mathbb R^{14}$ (enumerated in Ref.~\cite{oc_chep_proceeding}) as inputs and embeds them into a 256-dimensional space by a fully connected layer: $z^{(1)}_{i} \coloneqq W^{(1)} z^{(0)}_{i}$, with learnable weights $W^{(1)}\in\mathbb{R}^{256\times 14}$. 
Five subsequent layers with width 256, $\relu$ activations, and residual connections of the form $z^{(\ell+1)}_{i} \coloneqq \sqrt{\beta}\, W^{(\ell+1)} \relu\bigl(z^{(\ell)}_{i}\bigr) +  \sqrt{1-\beta}\, z^{(\ell)}_{i}$ (where $l=1,\dots,5$ and  $\beta=0.4$) are applied.
A final layer is applied to map the hidden representations $z_i^{(6)}$ down to an 8-dimensional space $h^{\text{GC}}_i\coloneqq W^{(7)}\relu(z_i^{(6)})$.

The network is trained with an attractive loss $L^\text{att}$ and a repulsive hinge loss $L^\text{rep}$:
\begin{align}
    L^\text{att} &\coloneqq \frac 1{|I_\text{att}|} \sum_{(i, j)\in I_\text{att}} \|h^{\text{GC}}_i - h^{\text{GC}}_j\|^2,
        \qquad I_\text{att}\coloneqq \{(i, j)\,|\, 1\leq i,j\leq N_\text{hits},\ \pi_i=\pi_j,\ \pi_i\ \text{of interest}\},\\
    L^\text{rep} &\coloneqq \frac 1{|I_\text{att}|} \sum_{(i, j)\in I_\text{rep}} \relu\bigl(1-\|h^{\text{GC}}_i - h^{\text{GC}}_j\|^2\bigr),
        \qquad I_\text{rep}\coloneqq \{(i, j)\,|\, 1\leq i,j\leq N_\text{hits},\ \pi_i\neq \pi_j\ \text{or noise}\},
        \notag
\end{align}
where $\pi_i$ denotes the particle of hit $i$ and \enquote{of interest} is defined as in~\autoref{sec:metrics}.
Note that these loss functions differ from those of Ref.~\cite{ju_exatrkx_2021} in how they deal with low-$\pt$ hits.
In addition, Ref.~\cite{ju_exatrkx_2021} includes all pairs of hits in non-consecutive detector layers in the repulsive potential.

\begin{figure}
    \centering
    \includegraphics[height=7cm]{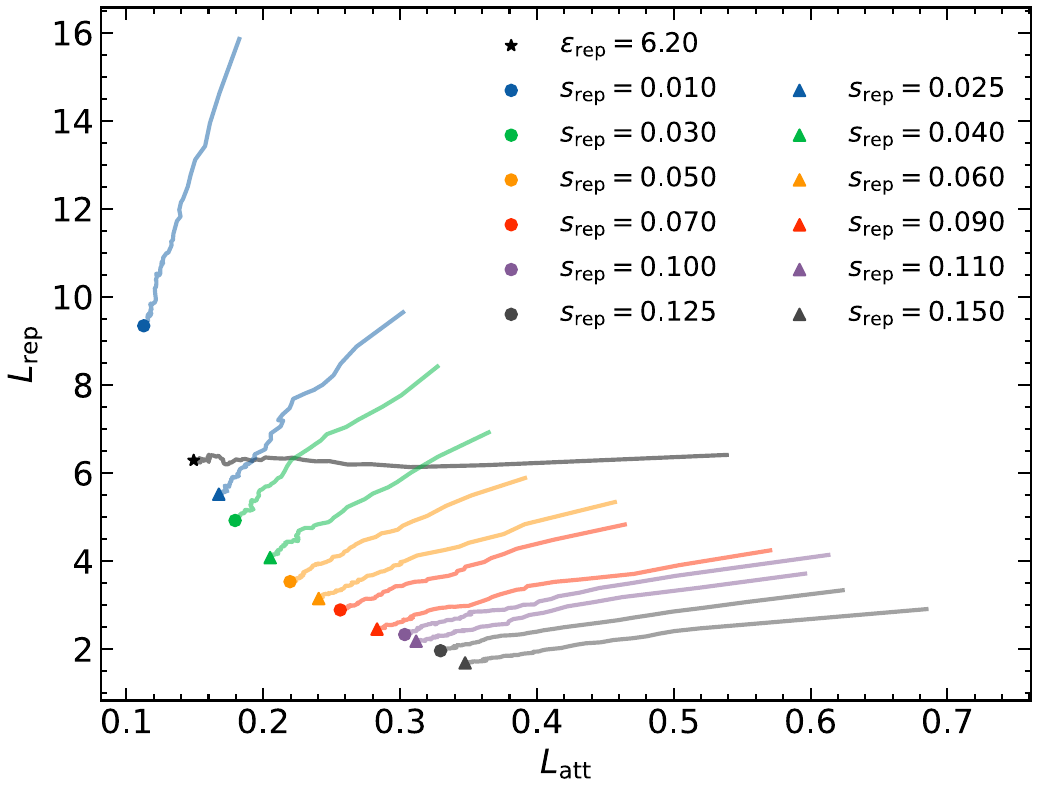}
    \caption{Identifying the Pareto front. Each colored line corresponds to the training trajectory of one model through $(L_\text{att}, L_\text{rep})$ space. Models with a specified value of $s_\text{rep}$ use linear scalarization of the two objectives, while the model with $\epsilon_\text{rep}$ uses MDMM (with $\epsilon_\text{rep}$ denoting the constraint on $L_\text{rep}$). }
    \label{fig:pareto}
\end{figure}

Balancing the attractive and repulsive loss strongly influences the performance of the graph construction. 
We find that linear scalarization of the two objectives, $L\coloneqq L^\text{att} + s_\text{rep}L^\text{rep}$ (we use $s_\text{rep}=0.06$), leads to overall good convergence with a corresponding convex Pareto front (\autoref{fig:pareto}).
In preparation for dealing with the many objective functions of the OC approach, we have furthermore investigated the use of the modified method of differential modifiers (MDMM)~\cite{NIPS1987_a87ff679} and optimized $L^\text{att}$ subject to a constraint on $L^\text{rep}$ (\autoref{fig:pareto}).
This achieved identical results but did not significantly simplify the optimization process (note that this might become more relevant when track parameter prediction is incorporated as an additional objective).

\begin{figure}
    \centering
    \includegraphics[height=6.5cm]{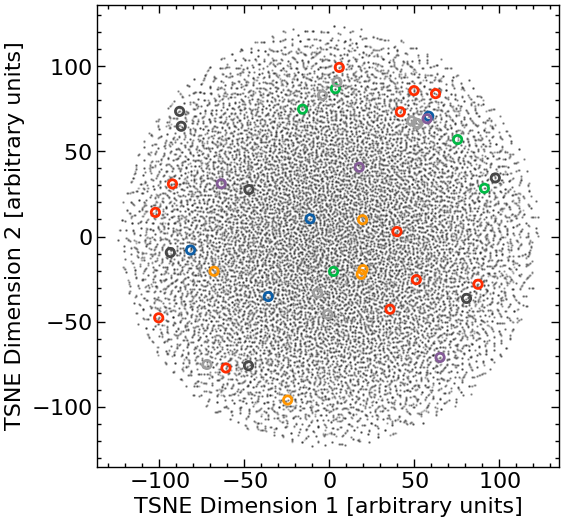}
    \quad\qquad
    \includegraphics[height=6.5cm]{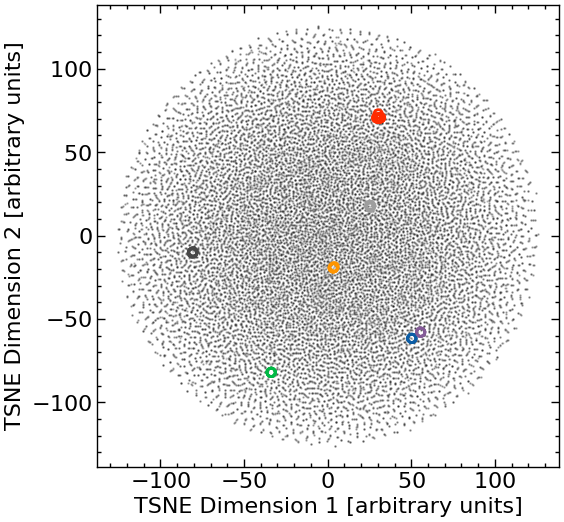}
    \caption{t-SNE projection of the input space (left) and the graph construction embedding space $h^{\text{GC}}$ (right). The hits of seven randomly selected particles of interest have been colored.}
    \label{fig:gc:tsne}
\end{figure}

A visual representation of track hits embedded in the learned clustering space is shown in~\autoref{fig:gc:tsne}.
While the embedding looks near-perfect to the eye, Figure~\autoref{fig:vs_eps:after_gc} (discussed in the next section) confirms that this initial clustering space is still insufficient to directly reconstruct tracks from.

\begin{figure}
    \centering
    \includegraphics[height=6.5cm]{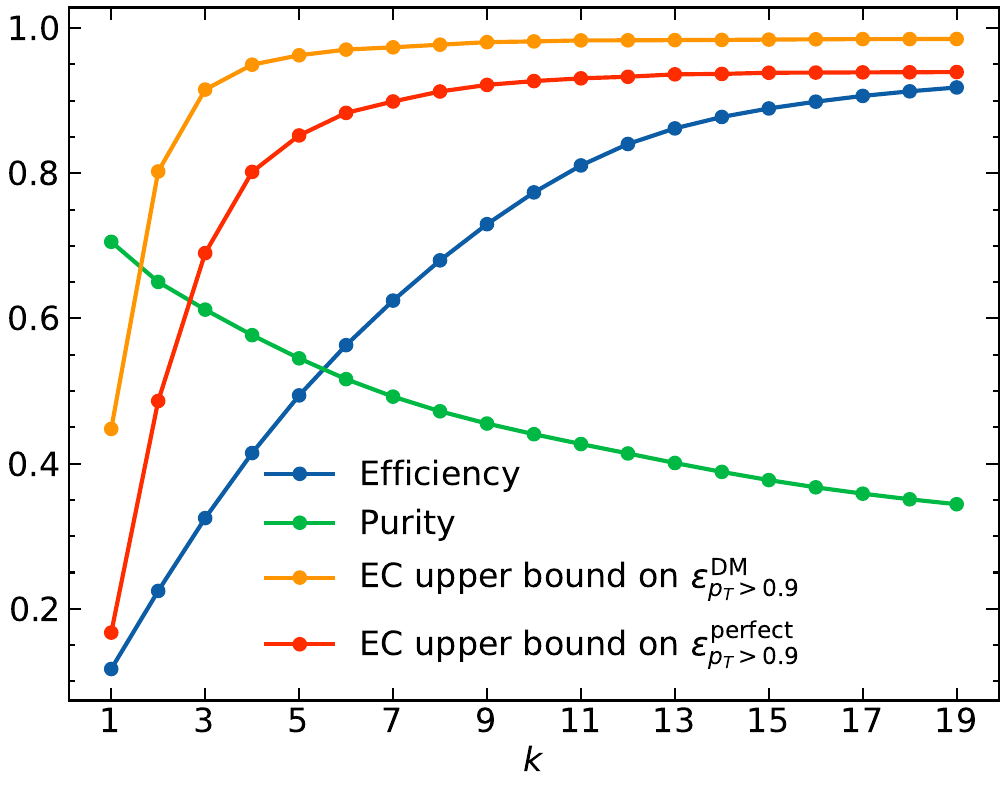}
    \caption{Choosing the number of nearest neighbors ($k$) for graph construction in our 2.0 pipeline. Note that efficiency and purity are relative to a fully connected graph of all hits per particle (rather than a path-graph of layer-to-layer connections). 
    }
    \label{fig:gc:choosing_k}
\end{figure}

The graph is then built from this latent space using $k$-nearest neighbors (kNN) while also limiting the maximal edge length to $1$ (arbitrary units).
To quantify the quality of the constructed graph, we call an edge \emph{true}, if it connects two hits of the same particle (any edge that connects to a noise hits is \emph{false}). 
We call an edge \emph{of interest} if at least one of the two hits belongs to a particle of interest.
The quality of the constructed graph vs $k$ is quantified in~\autoref{fig:gc:choosing_k} in terms of four metrics:
\begin{itemize}
    \item 
    \textbf{Efficiency:} The number of true edges of interest normalized to the number of possible true edges of interest ($\sum_{\pi\ \text{of interest}}{N_\text{hits}^\pi \choose 2}$, where $N_\text{hits}^\pi$ is the number of hits for a particle $\pi$)
    \item 
    \textbf{Purity:} The number of true edges of interest normalized to the number of edges of interest.
    \item 
    \textbf{Upper bounds on figures of merit:} 
    As introduced in Ref.~\cite{oc_chep_proceeding}, we define upper bounds of an EC pipeline for $\effDmPt{0.9}$ and $\effPerfectPt{0.9}$ by calculating both metrics for tracks given by the connected components of the edge-subgraph that only contains true edges.
    It was shown that these upper bounds do not necessarily hold for an OC pipeline, but they are still useful in quantifying the connectivity of the graph in a way that relates to the key tracking metrics.
\end{itemize}
While the overall efficiency of the GC step flattens out slowly, the upper bounds on $\effDmPt{0.9}$ and $\effPerfectPt{0.9}$ show only very limited increases for $k>10$. 
We therefore choose $k=10$ for the remainder of this section.
At this point, an average of $468\times10^3$ edges are built with an efficiency of $77\%$, a purity of $44\%$, and EC upper bounds on $\effDmPt{0.9}$ and $\effPerfectPt{0.9}$ of $98\%$ and $92\%$.

\subsection{Object condensation}
The GNN used to perform OC is identical to that in Ref.~\cite{oc_chep_proceeding} except for different input shapes and an additional residual connection that connects the 8-dimensional embedding space ($h_i^\mathrm{GC}$) that was used for GC with the final OC output coordinates.

The model is built from \emph{interaction network} layers~\cite{battaglia_interaction_2016} with residual connections in the node updates.
The node features $x_i^{(0)}$ are the input features $z_i^{(0)}\in \mathbb R^{14}$ described in~\autoref{sec:dataset} concatenated with the GC embedding $h^\text{GC}_i\in\mathbb R^8$, i.e., $x_i^{(0)}\coloneqq [z_i^{(0)}, h^\text{GC}_i]$. 
The edge features $e_{ij}^{(0)}$ ($(i, j)\in I$, where $I$ represents the edges of the graph) are given by $e_{ij}^{(0)}\coloneqq [x_i^{(0)}-x_j^{(0)},x_i^{(0)}+x_j^{(0)}]\in\mathbb R^{44}$.

Node and edge features are first encoded, $x_i^{(1)} \coloneqq W^\text{enc}_\text{node}x_i^{(0)}$, $e_{ij}^{(1)} \coloneqq W^\text{enc}_\text{edge}e_{ij}^{(0)}$, $W^\text{enc}_\text{node}\in\mathbb R^{192\times 22}$, and $W^\text{enc}_\text{edge}\in\mathbb R^{192\times 44}$. 
Then, five iterations of message passing are performed with 
\begin{align}
\begin{split}
    e_{ij}^{(k+1)} &\coloneqq \bigl(\Phi^{(k+1)}\circ \relu\bigr)\Big(\bigl[x_i^{(k)}, x_j^{(k)}, e_{ij}^{(k)}\bigr]\Big),\\
    x_i^{(k+1)} &\coloneqq \sqrt\beta\, \Psi^{(k+1)}\Bigl(\bigl[\relu\bigl(x_{i}^{(k)}\bigr), {\textstyle\sum\nolimits_{j\in\mathcal N_{i}}} e_{ij}^{(k+1)}\bigr]\Bigr) + \sqrt{1-\beta}\, x_i^{(k)}.
    \label{eq:resin}
\end{split}
\end{align}
Here, $(i,j)\in I$, $k=1,\dots,5$, and $\mathcal N_i$ denotes the neighborhood of $i$. $\Phi$ and $\Psi$ are FCNNs with $\relu$ activations, a layer width of 192, and one hidden layer.
$\beta$ has been chosen to be $0.5$.
Finally, the clustering coordinates are decoded (and added to a residual from the GC space) as
\begin{align}
    c_i \coloneqq 
    \sqrt{\beta'}\, W^\text{dec}_{\text{c}} \relu(x_i^{(6)}) + \sqrt{1-\beta'}\, [h_i^{\text{GC}}, 0, \dots, 0],
\end{align}
where we chose $\beta'=0.5$, $W_\text{c}^\text{dec}\in \mathbb R^{24\times 192}$, and $h_i^\text{GC}$ is zero-padded to $\mathbb R^{24}$. 
The condensation likelihoods are decoded as $\beta_i \coloneqq \sigma\bigl(W^\text{dec}_\beta \relu(x_i^{(6)})\bigr)$, where $\sigma$ is the logistic function, and $W^\text{dec}_\beta\in \mathbb R^{1\times 192}$.  
The total number of parameters of this model is $1.9\times 10^6$. 

To train the model, we use the object condensation loss functions~\cite{Kieseler_condensation_2020} as described in Ref.~\cite{oc_chep_proceeding} with the following hyperparameters: $s_\beta=0$, $s_\text{rep}=0.74$, and $q_\text{min}=0.01$.
\subsection{Post-processing and results}
\begin{figure}
    \centering
    \subfloat[]{
        \includegraphics[height=6cm]{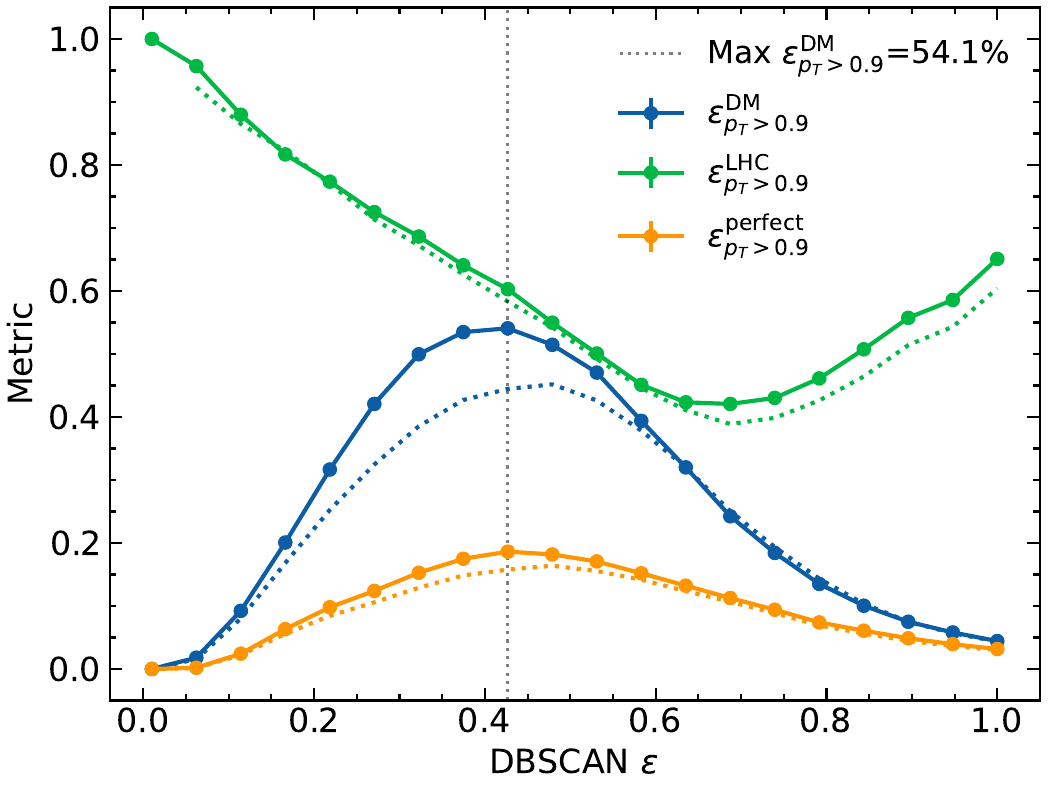}
        \label{fig:vs_eps:after_gc}
    }
    \subfloat[]{
        \includegraphics[height=6cm]{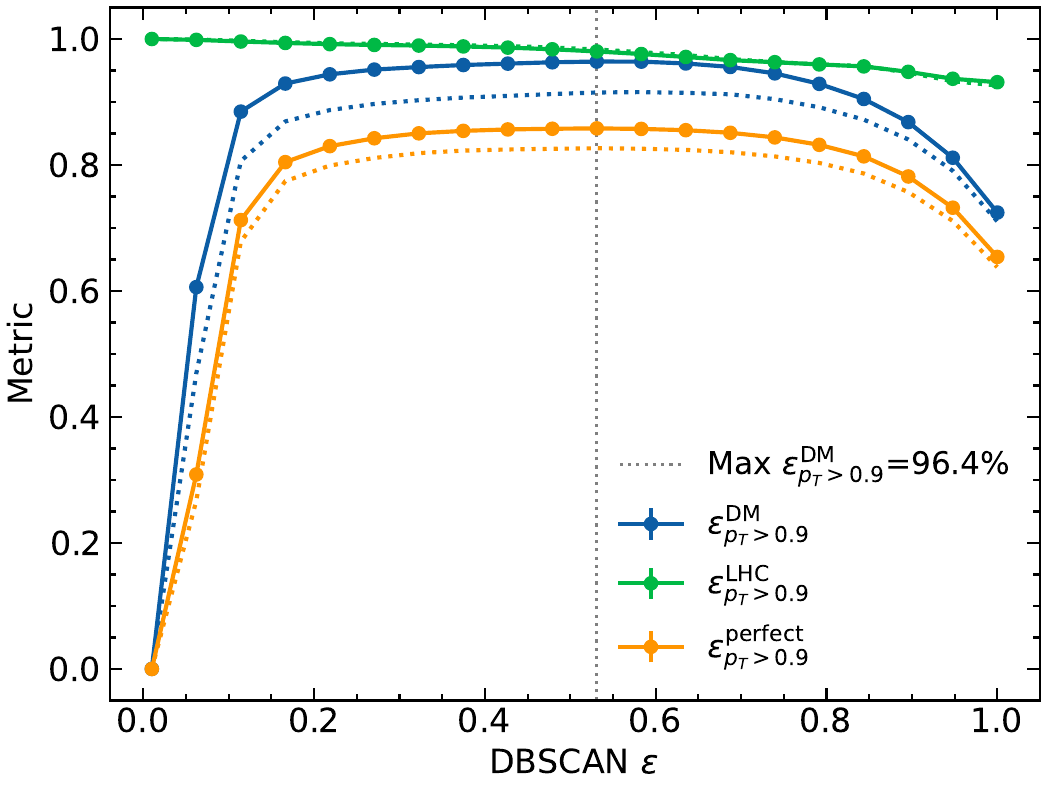}
        \label{fig:vs_eps:after_oc}
    }
    \caption{
        Rendering tracks from the learned clustering space with DBSCAN. 
        The left plot shows track reconstruction using the embedding space $h^{\text{GC}}$ used for graph construction, and the right plot shows the significantly improved tracking performance using the OC embedding space.
        Dashed lines correspond to DBSCAN with $k=4$.
    }
\end{figure}

\begin{figure}
    \centering
    \includegraphics[height=6cm]{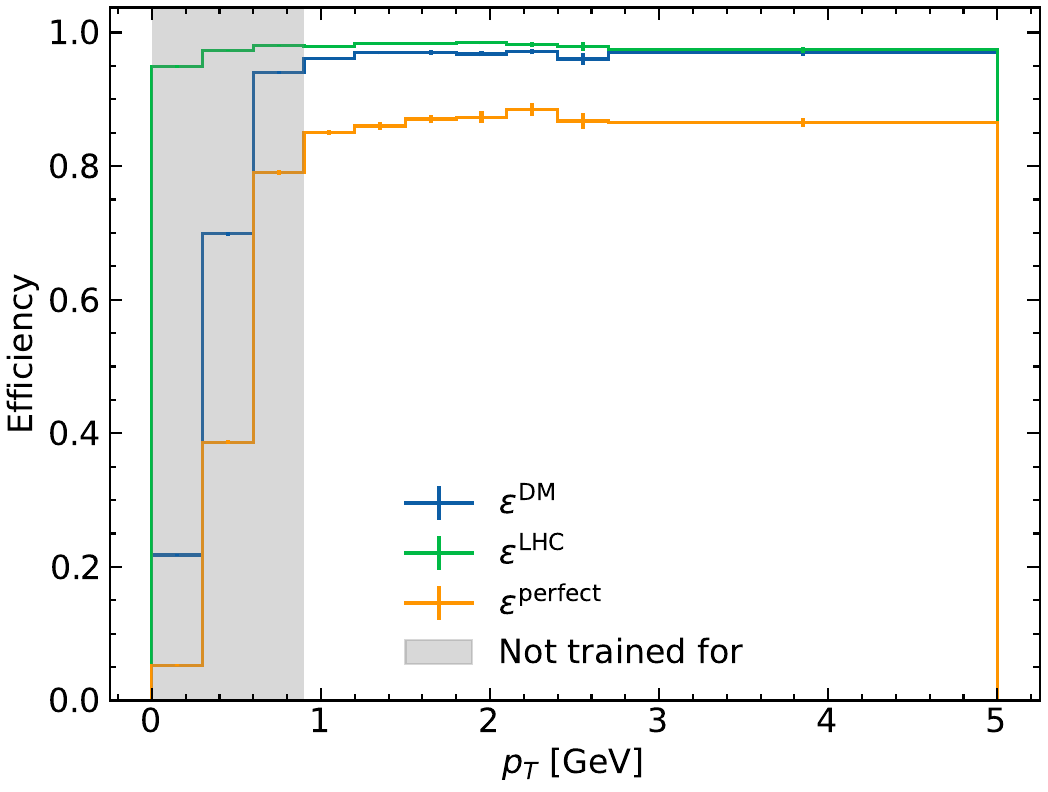}
    \includegraphics[height=6cm]{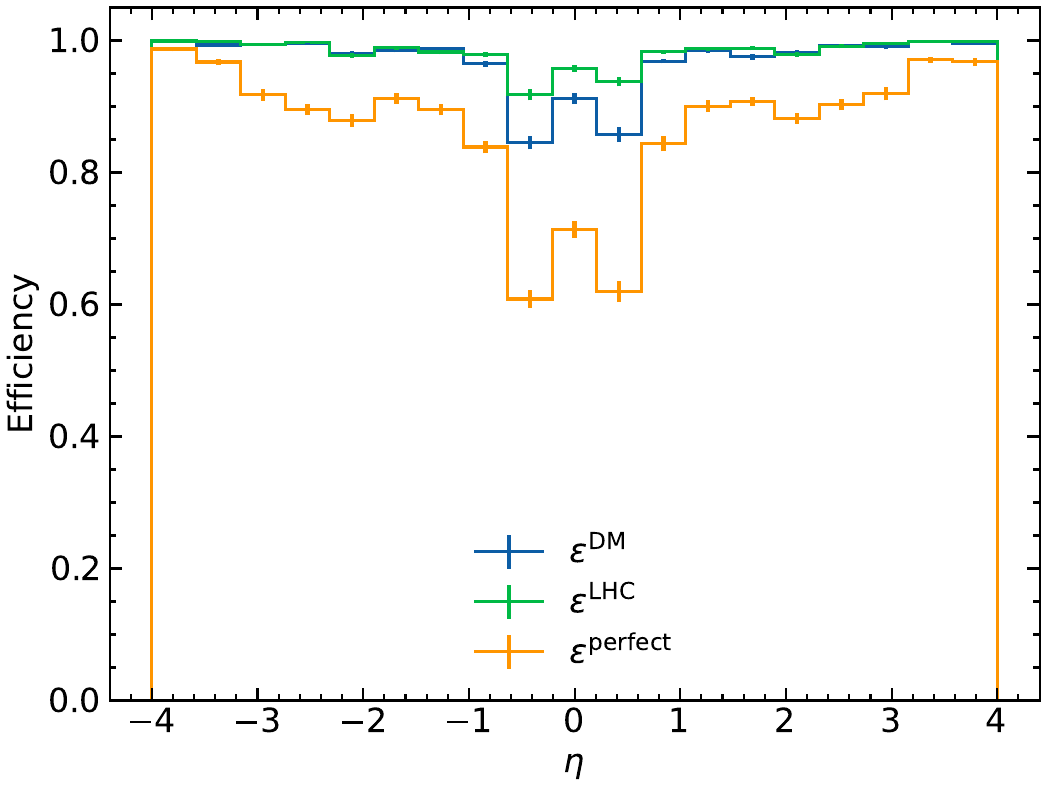}
    \includegraphics[height=6cm]{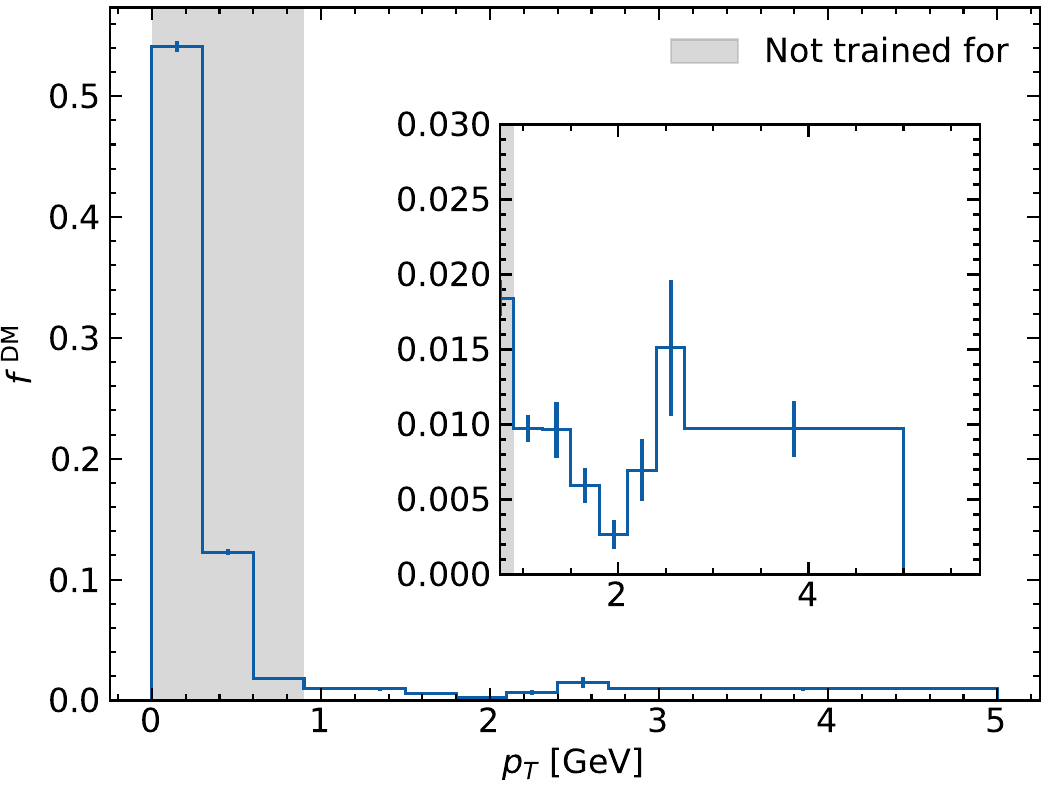}
    \includegraphics[height=6cm]{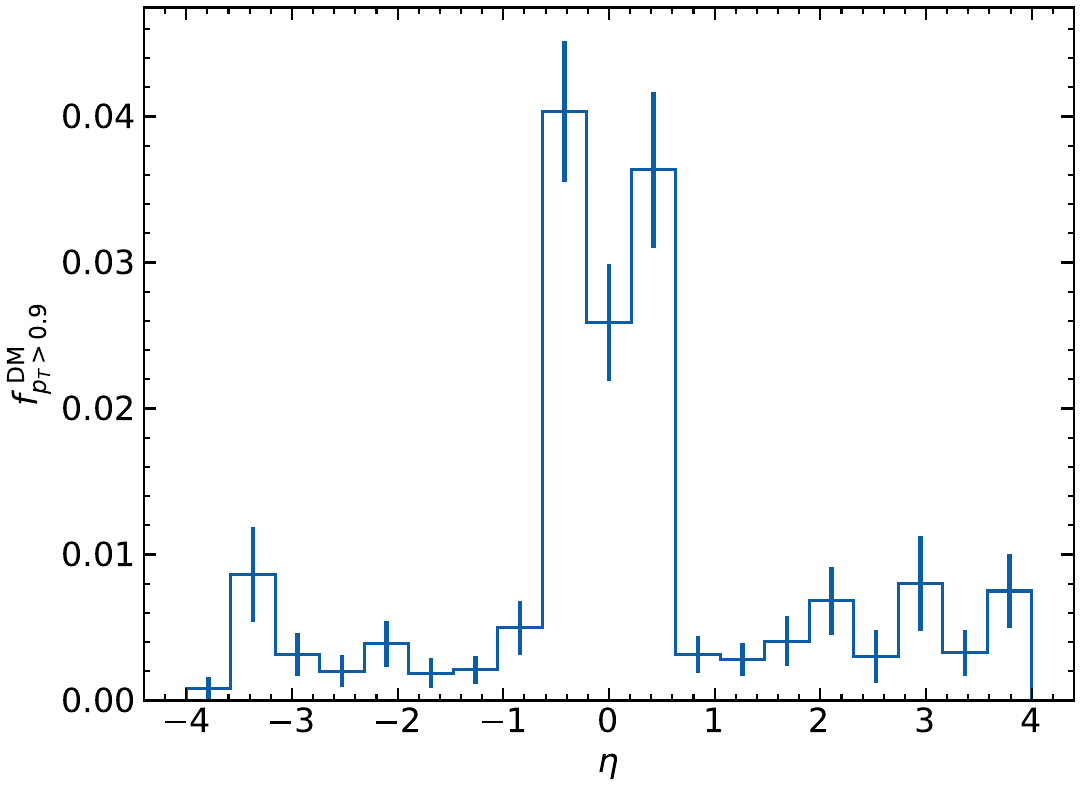}
    \caption{Tracking performance in bins of $\pt$ and $\eta$.}
    \label{fig:oc_performance_pt_eta}
\end{figure}

We use Density-Based Spatial Clustering of Applications with Noise (DBSCAN)~\cite{ester1996density} to determine clusters in the OC clustering space $c$. 
DBSCAN is an iterative clustering algorithm with two parameters, $\epsilon$ (the size of the neighborhood of a point that is considered when merging clusters), and $k$ (minimum number of points within a neighborhood for the points to be considered a \emph{core point}). 
In our case, $k=1,2,3$ produces equal results (this is related to the fact that clusters with less than three hits are discarded as track candidates) and $k\geq 4$ degrades performance (see Figure~\autoref{fig:vs_eps:after_oc}). 
Therefore, all results use $k=1$.
Based on Figure~\autoref{fig:vs_eps:after_oc}, we choose $\epsilon=0.53$. 
The broad plateau in $\epsilon$ vs $\effDmPt{0.9}$ shows that the clusters are well separated from each other and from noise. 
We obtain $\effDmPt{0.9}=96.4\%$, $\effLhcPt{0.9}=98.0\%$, $\effPerfectPt{0.9}=85.8\%$ and $f_{\pt>0.9}=0.9\%$.
These metrics uniformly improve our previous results in Ref.~\cite{oc_chep_proceeding}.
All metrics are presented vs $\pt$ and vs $\eta$ in~\autoref{fig:oc_performance_pt_eta}.


\section{Progress on other architectures}
\label{sec:other_architectures}
While the model presented in the main part of this paper is built on interaction networks, many other architectures are left to be explored in detail.
For example, \emph{GravNet}~\cite{Qasim:2019otl} layers have been successfully used with an OC approach for multi-particle reconstruction in high occupancy imaging
calorimeters~\cite{qasim_multi-particle_2021,qasim2022endtoend}. 
A similar architecture is also being investigated for the Belle II outer tracker, motivating studies for its suitability for an HL-LHC application.
In our first experiments, we confirmed that an architecture of four (slightly modified) GravNet layers can achieve $\effDmPt{0.9}>90\%$ using $k=4$ kNN with 8-dimensional latent-spaces. 
However, a careful optimization of hyperparameters is needed for an apples-to-apples comparison with the pipeline described in~\autoref{sec:pipeline2.0}.

Shifting from a very edge-oriented perspective on tracking models (e.g., EC approaches) to one that emphasizes embeddings (e.g. OC) also motivates to consider non-GNN architectures.
The success of transformer architectures in encoding complex relationships between tokens and a large body of work related to hardware optimization makes this class of models attractive for the tracking problem.
A key challenge for the application to tracking is to find efficient realizations of the attention mechanism in order to overcome the quadratic scaling with the attention window size (that would need to encompass all hits of an event in the naive scaled dot product attention).
In our approach, this is achieved by using relative positional encoding and Exact Euclidean Locality Sensitive Hashing~\cite{data04_e2lsh} (E2LSH) with an $\mathcal O(n\log n)$ scaling.
The resulting model features more regular and parallelizable computations than GNN models and avoids the performance bottleneck of GPU-implementations of kNN graph building. 
The model is trained with contrastive learning loss functions and hard negative mining.
While we are still working on detailed evaluations in terms of tracking performance, we observe up to hundred-fold inference speedups on a Quadro RTX 6000 compared to the pipeline presented in~\autoref{sec:pipeline2.0}.

\section{Conclusions}
This paper presents an object condensation approach to charged particle tracking at the worst-case pileup conditions expected at the HL-LHC.
Replacing the geometric graph construction stage of our previous pipeline~\cite{oc_chep_proceeding} with an adaptation of the metric learning approach from Ref.~\cite{ju_exatrkx_2021} allowed us to make our pipeline more streamlined while significantly improving tracking performance on the pixel detector of the TrackML dataset. 
Ongoing work to lower memory consumption will make it possible to apply this pipeline to the full detector.
Our pipeline is part of our open-source project \cite{Lieret_gnn_tracking_An_open-source} that implements various OC tracking architectures in a modular and extensible Python package. 

In addition, we discussed ongoing research into other architectures, such as GravNet and sparse transformers.
The first results with sparse transformers show major improvements in inference time due to hardware-friendly implementations.
%

\Acknowledgements
We would like to thank Javier Duarte, Peter Elmer, Philip Chang, Jonathan Guiang, David Lange, and Savannah Thais for helpful discussions and support.
%
%
%
%
%



%
\printbibliography

@InProceedings{amrouche_tracking_2020,
editor="Escalera, Sergio and Herbrich, Ralf",
booktitle="The NeurIPS '18 Competition",
publisher="Springer International Publishing",
address="Cham",
pages="231--264",
	title = {The {Tracking} {Machine} {Learning} challenge : {Accuracy} phase},
	shorttitle = {The {Tracking} {Machine} {Learning} challenge},
	urldate = {2022-01-10},
	journal = {arXiv:1904.06778 [hep-ex, physics:physics]},
	author = {Amrouche, Sabrina and Basara, Laurent and Calafiura, Paolo and Estrade, Victor and Farrell, Steven and Ferreira, Diogo R. and Finnie, Liam and Finnie, Nicole and Germain, Cécile and Gligorov, Vladimir Vava and Golling, Tobias and Gorbunov, Sergey and Gray, Heather and Guyon, Isabelle and Hushchyn, Mikhail and Innocente, Vincenzo and Kiehn, Moritz and Moyse, Edward and Puget, Jean-Francois and Reina, Yuval and Rousseau, David and Salzburger, Andreas and Ustyuzhanin, Andrey and Vlimant, Jean-Roch and Wind, Johan Sokrates and Xylouris, Trian and Yilmaz, Yetkin},
	year = {2020},
 eprint={https://doi.org/10.1007/978-3-030-29135-8{\_}9},
}

@article{amrouche_tracking_2021,
	title = {The {Tracking} {Machine} {Learning} challenge : {Throughput} phase},
	shorttitle = {The {Tracking} {Machine} {Learning} challenge},
	url = {http://arxiv.org/abs/2105.01160},
	urldate = {2022-01-10},
	journal = {arXiv:2105.01160 [cs.LG]},
	author = {Amrouche, Sabrina and Basara, Laurent and Calafiura, Paolo and Emeliyanov, Dmitry and Estrade, Victor and Farrell, Steven and Germain, Cécile and Gligorov, Vladimir Vava and Golling, Tobias and Gorbunov, Sergey and Gray, Heather and Guyon, Isabelle and Hushchyn, Mikhail and Innocente, Vincenzo and Kiehn, Moritz and Kunze, Marcel and Moyse, Edward and Rousseau, David and Salzburger, Andreas and Ustyuzhanin, Andrey and Vlimant, Jean-Roch},
	month = may,
	year = {2021},
    note={submitted to Computing and Software for Big Science},
    eprint={https://doi.org/10.48550/arXiv.2105.01160},
}

@article{battaglia_interaction_2016,
      title="{Interaction Networks for Learning about Objects, Relations and Physics}", 
      author={Peter W. Battaglia and Razvan Pascanu and Matthew Lai and Danilo Rezende and Koray Kavukcuoglu},
      year={2016},
      eprint={1612.00222},
      archivePrefix={arXiv},
      primaryClass={cs.AI},
      note={Preprint at \url{https://arxiv.org/abs/1612.00222}}
}

@article{Cerati_2016,
	doi = {10.1051/epjconf/201612700010},
  
	url = {https://doi.org/10.1051%2Fepjconf%2F201612700010},
  
	year = 2016,
	publisher = {{EDP} Sciences},
  
	volume = {127},
  
	pages = {00010},
  
	author = {Giuseppe Cerati and Peter Elmer and Slava Krutelyov and Steven Lantz and Matthieu Lefebvre and Kevin McDermott and Daniel Riley and Matev{\v{z}
} Tadel and Peter Wittich and Frank Würthwein and Avi Yagil},
  
	editor = {R. Frühwirth and E. Brondolin and B. Kolbinger and W. Waltenberger},
  
	title = {Kalman Filter Tracking on Parallel Architectures},
  
	journal = {{EPJ} Web of Conferences}
}

@inproceedings{data04_e2lsh,
author = {Datar, Mayur and Immorlica, Nicole and Indyk, Piotr and Mirrokni, Vahab S.},
title = {Locality-Sensitive Hashing Scheme Based on p-Stable Distributions},
year = {2004},
isbn = {1581138857},
publisher = {Association for Computing Machinery},
address = {New York, NY, USA},
url = {https://doi.org/10.1145/997817.997857},
doi = {10.1145/997817.997857},
booktitle = {Proceedings of the Twentieth Annual Symposium on Computational Geometry},
pages = {253–262},
numpages = {10},
location = {Brooklyn, New York, USA},
series = {SCG '04}
}

@inproceedings{ester1996density,
  title={A density-based algorithm for discovering clusters in large spatial databases with noise.},
  author={Ester, Martin and Kriegel, Hans-Peter and Sander, Jorg and Xu, Xiaowei and others},
  booktitle={kdd},
  volume={96},
  number={34},
  pages={226--231},
  year={1996}
}

@article{ju_exatrkx_2021,
	title = {Performance of a geometric deep learning pipeline for {HL}-{LHC} particle tracking},
	volume = {81},
	issn = {1434-6044, 1434-6052},
	url = {https://link.springer.com/10.1140/epjc/s10052-021-09675-8},
	eprint = {htttps://doi.org/10.1140/epjc/s10052-021-09675-8},
	language = {en},
	number = {10},
	urldate = {2022-01-10},
	journal = {The European Physical Journal C},
	author = {Ju, Xiangyang and Murnane, Daniel and Calafiura, Paolo and Choma, Nicholas and Conlon, Sean and Farrell, Steven and Xu, Yaoyuan and Spiropulu, Maria and Vlimant, Jean-Roch and Aurisano, Adam and Hewes, Jeremy and Cerati, Giuseppe and Gray, Lindsey and Klijnsma, Thomas and Kowalkowski, Jim and Atkinson, Markus and Neubauer, Mark and DeZoort, Gage and Thais, Savannah and Chauhan, Aditi and Schuy, Alex and Hsu, Shih-Chieh and Ballow, Alex and Lazar, Alina},
	month = oct,
	year = {2021},
	pages = {876},
}

@article{Kieseler_condensation_2020,
	doi = {10.1140/epjc/s10052-020-08461-2},
	url = {https://doi.org/10.1140\%2Fepjc%2Fs10052-020-08461-2},
	year = 2020,
	month = {sep},
	publisher = {Springer Science and Business Media {LLC}
},
	volume = {80},
	number = {9},
	author = {Jan Kieseler},
	title = {Object condensation: one-stage grid-free multi-object reconstruction in physics detectors, graph, and image data},
  
	journal = {The European Physical Journal C}
}

@misc{oc_chep_proceeding,
      title={An Object Condensation Pipeline for Charged Particle Tracking at the High Luminosity LHC}, 
      author={Kilian Lieret and Gage DeZoort},
      year={2023},
      eprint={2309.16754},
      archivePrefix={arXiv},
      primaryClass={physics.data-an}
}

@misc{Lieret_gnn_tracking_An_open-source,
    author = {Lieret, Kilian and DeZoort, Gage},
    title = {{gnn{\_}tracking: An open-source GNN tracking project}},
    howpublished = {\url{https://github.com/gnn-tracking/gnn_tracking}}
}

@article{lazar_accelerating_2023,
	doi = {10.1088/1742-6596/2438/1/012008},
	url = {https://doi.org/10.1088\%2F1742-6596\%2F2438\%2F1\%2F012008},
	year = 2023,
	month = {feb},
	publisher = {{IOP} Publishing},
	volume = {2438},
	number = {1},
	pages = {012008},
	author = {Alina Lazar and Xiangyang Ju and Daniel Murnane and Paolo Calafiura and Steven Farrell and Yaoyuan Xu and Maria Spiropulu and Jean-Roch Vlimant and Giuseppe Cerati and Lindsey Gray and Thomas Klijnsma and Jim Kowalkowski and Markus Atkinson and Mark Neubauer and Gage DeZoort and Savannah Thais and Shih-Chieh Hsu and Adam Aurisano and Jeremy Hewes and Alexandra Ballow and Nirajan Acharya and Chun-yi Wang and Emma Liu and Alberto Lucas},
	title = {Accelerating the Inference of the Exa.{TrkX} Pipeline},
	journal = {Journal of Physics: Conference Series}
}

@inproceedings{NIPS1987_a87ff679,
 author = {Platt, John and Barr, Alan},
 booktitle = {Neural Information Processing Systems},
 editor = {D. Anderson},
 pages = {},
 publisher = {American Institute of Physics},
 title = {Constrained Differential Optimization},
 url = {https://proceedings.neurips.cc/paper_files/paper/1987/file/a87ff679a2f3e71d9181a67b7542122c-Paper.pdf},
 volume = {0},
 year = {1987}
}

@article{Qasim:2019otl,
    author = "Qasim, Shah Rukh and Kieseler, Jan and Iiyama, Yutaro and Pierini, Maurizio",
    title = "{Learning representations of irregular particle-detector geometry with distance-weighted graph networks}",
    eprint = "1902.07987",
    archivePrefix = "arXiv",
    primaryClass = "physics.data-an",
    doi = "10.1140/epjc/s10052-019-7113-9",
    journal = "Eur. Phys. J. C",
    volume = "79",
    number = "7",
    pages = "608",
    year = "2019"
}

@article{qasim_multi-particle_2021,
	title = {Multi-particle reconstruction in the {High} {Granularity} {Calorimeter} using object condensation and graph neural networks},
	url = {http://arxiv.org/abs/2106.01832},
	urldate = {2022-01-10},
	journal = {arXiv:2106.01832 [physics.ins-det]},
	author = {Qasim, Shah Rukh and Long, Kenneth and Kieseler, Jan and Pierini, Maurizio and Nawaz, Raheel},
	month = jun,
	year = {2021},
    note={\url{https://doi.org/10.48550/arXiv.2106.01832}},
}

@article{qasim2022endtoend,
    author = "Qasim, Shah Rukh and Chernyavskaya, Nadezda and Kieseler, Jan and Long, Kenneth and Viazlo, Oleksandr and Pierini, Maurizio and Nawaz, Raheel",
    title = "{End-to-end multi-particle reconstruction in high occupancy imaging calorimeters with graph neural networks}",
    archivePrefix = "arXiv",
    primaryClass = "physics.ins-det",
    journal = "Eur. Phys. J. C",
    volume = "82",
    number = "8",
    pages = "753",
    year = "2022",
    eprint={https://doi.org/10.1140/epjc/s10052-022-10665-7},
}
\end{document}